# Experimental Characterization of Magnetic Materials for the Magnetic Shielding of Cryomodules in Particle Accelerators

*Sanjay SAH[1], Ganapati MYNENI[2], Jayasimha ATULASIMHA[1]*

[1] *Virginia Commonwealth University, Dept. of Mechanical and Nuclear Engineering,*

*401 W. Main Street, Richmond, VA 23284-3015*

[2]*Thomas Jefferson National Accelerator Facility*

*12000 Jefferson Avenue, Newport News, VA 23606*

## ABSTRACT

*The magnetic properties of two important passive magnetic shielding materials (A4K and Amumetal) for accelerator applications, subjected to various processing and heat treatment conditions are studied comprehensively over a wide range of temperatures: from Cryogenic to room temperature. We analyze the effect of processing on the extent of degradation of the magnetic properties of both materials and investigate the possibility of restoring these properties by re-annealing.*

## I. INTRODUCTION

Magnetic shielding is extremely vital for the enhanced performance of cryomodules (CMs) of particle accelerators. This can be understood in terms of the effect of stray magnetic fields on the quality factor ($Q_o$) of the CM that is given by [1]

$$Q_o = \frac{G}{R_s}, \quad (1)$$

Here, G is the geometric factor of the accelerating cavity and $R_s$ is the cavity surface resistance. The cavity surface resistance ($R_s$) can be divided into contributions from the surface magnetic field ($R_H$) and other components ($R_{other}$). The $R_H$ can be estimated using the equation (2) as follows [1]

$$R_H = \frac{H_{ext}}{2H_{c2}} R_n \approx 9.49 \times 10^{-12} H_{ext}\sqrt{f}, \quad (2)$$

$H_{ext}$ is the external field that in this case is the earth's magnetic field (~500 mG), f is the fundamental frequency of the Niobium cavity, $H_{c2}$ is the type-II superconductor (Niobium) magnetic quench field and $R_n$ is the normal conducting resistance of niobium. Thus, it is clear that a high stray magnetic field increases the cavity surface resistance, thereby degrading the cavity's quality factor. Furthermore, during quenching, of the cryomodule cavities, the Nb is not in its superconducting state and therefore magnetic flux can penetrate the cavity.

These issues can be effectively addressed by the appropriate use of magnetic shields [2] that reduce the magnetic field in a prescribed region. The magnetic shielding can be provided by an active shield [3] that uses a magnetic field produced by utilizing a superconducting coil to cancel an external magnetic field or a passive shield [4] that works by drawing the field onto itself, providing a path for the field lines around the shielding volume and minimizing the magnetic field inside the cryomodule.

Here we study magnetic properties of materials used in passive shields that mitigate the effect of the Earth's axial and transverse magnetic field components on cryomodules. Specifically, we focus on understanding of the manner in which magnetic permeability varies with temperature, applied deformation during manufacturing and heat treatment. While some prior work exists on characterizing the magnetic [2], [4-6], [8] properties, a comprehensive study of the effect of deformation during the manufacturing process and annealing on the magnetic permeability of shielding materials over a broad range of temperatures (cryogenic to room temperature) is not available. This paper bridges this gap in knowledge by performing such experimental studies on these magnetic materials.

The current materials of interest for magnetic shielding are Amumetal and A4K and therefore these materials are studied in this paper. Both materials are high nickel content alloys. A4K is composed of 81% nickel, 4.5% molybdenum and rest iron by weight. Amumetal is composed of 80% nickel, 4.5% molybdenum and rest iron by weight. The samples studies were obtained from Amuneal Manufacturing Corporation [5].

## II. EXPERIMENTAL METHODS

*i. Sample Preparation*

Two mill-annealed samples of A4K and Amumetal were obtained from Amuneal Manufacturing Corporation with planar dimensions 3'x3' and 1 mm in thickness (Fig 1 a). The samples were cut into 2 mm x 2 mm pieces of thickness 1mm (Fig 1 b) using Wire Electrical Discharge Machine (Wire-EDM) at the Jefferson Lab. The Wire-EDM was used so that the external stress induced in the samples during cutting is minimized. The magnetic properties of both un-annealed samples and those that were hydrogen annealed (pure hydrogen and dry atmosphere) at Amuneal Manufacturing Corporation at 1150°C for four hours were studied. We note that after the anneal process, the cooling rates for Amumetal and A4K were 200°C/h and 50°C/h respectively.

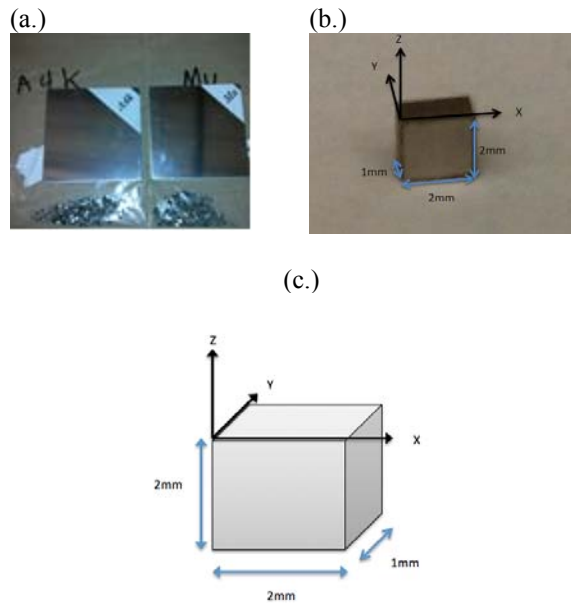

**FIG. 1.** (a.) Amumetal and A4K samples (b.) Amumetal or A4K sample after EDM cutting, and (c.) Schematic of Amumetal or A4K Sample with dimension.

Two samples of each metal were then deformed by applying bending stress, which is equivalent to a maximum tensile/compressive stress of 3.18 MPa. The deformation process is designed to produce the typical stress induced in samples by the manufacturing processes while fabricating the shields. The magnetic properties of the deformed samples were studied to understand the effect of this manufacturing process on permeability.

Finally, these deformed samples were annealed again and tested to determine if the magnetic properties that were degraded during the deformation process could be restored by appropriate heat treatment.

*ii. Magnetic Testing*

The magnetic characterization on the different samples (unannealed, annealed and deformed) at 50K, 100K, 150K, 200K, 250K and 300K was performed using a Quantum Design Versalab Vibrating Sample Magnetometer (VSM) at the Nanomaterial Core Characterization (NCC) Facility of the Virginia Commonwealth University (VCU). The magnetic characterization of samples annealed after deformation were tested only at 300K as explained later.

Magnetic moment as a function of field applied in the Z-direction (axes shown in Fig. 1 c) was collected for each sample at the different temperatures mentioned above. A SQUID (Quantum Design Magnetic Property Measurement System-3) magnetometer at the University of Maryland was used to obtain magnetic moment vs. applied field at temperature of 5K.

*iii. Demagnetizing Factor*

A demagnetizing field is generated when samples are magnetized. This needs to be correctly accounted for while reporting the magnetic moment at a given applied field. The effective field inside the sample that produces the moment is given as [7], [9-10],

$$ \tag{3} $$

Where, $H_{in}$ is the effective magnetic field inside the sample, $H_{app}$ is the applied magnetic field, N is demagnetizing factor that is influenced by the geometry of the sample and M is the magnetic moment.

N is approximately determined from the experimental data using $N \approx H_{app}/M$ from the linear region of M-H curve where $\chi$ is very large. (Details of the derivation and when this approximation holds can be found in Appendix -1 of this paper).

## III. RESULTS AND ANALYSIS

Regular, annealed and deformed samples of Amumetal and A4K were tested at the temperatures of 5K, 50K, 100K, 150K, 200K, 250K and 300K. The plots Fig. 2-4 respectively show the M-H curves of AMU metal without annealing, after annealing and after deformation, while Fig. 5-7 respectively show the M-H curves of A4K for the same conditions.

In both materials, irrespective of the processing condition, we note that saturation magnetization (the plots we show are zoomed and $M_s$ is not exactly researched at the highest field shown on the plot ~ $2\times10^4$ A/m, but the trends still stay the same) decreases with the increase in temperature as expected in any second order system. Also, as expected, deformed samples have the lowest permeability and need high fields to drive them to saturation due to the large number of defects that act as pinning sites and impede the magnetization rotation or movement of magnetic domains walls. The undeformed but unannealed samples show higher permeability, likely due to lesser defect density while the annealed samples show the best permeability as the annealing process greatly reduces the defects/pinning sites[11-13].

The comparative value of the low field permeability (differential permeability at 0.5 Oe, approximate magnitude of the Earth's magnetic field) and the intermediate field permeability (differential permeability at ~250 Oe and ~500 Oe) for two materials (Amumetal and A4K) are tabled at two temperatures: 5K and 300K (in Table.1 and Table. 2.). These temperatures are of relevance to the inner magnetic shield at cryogenic temperature and the outer magnetic field at room temperature respectively. In addition to confirming that permeability at both temperatures is highest for annealed samples and lowest for deformed samples at low fields, it also shows that the low field permeability of annealed Amumetal and A4K are comparable at 300 K while that of annealed A4K is significantly better than that of annealed Amumetal at Cryogenic temperature (5K). This suggests that A4K is better suited for shielding Earth's magnetic field at low temperatures and should be the preferred material for design of inner shields.

**Table. 1.** Permeability at 5K

| Material | Permeability ($\mu_r$) at 5K | | | |
|---|---|---|---|---|
| | $\mu_r=\Delta B/\Delta H$ at 0.5Oe (~40 A/m) | $\mu_r=B/H$ at ~250 Oe (~2×10$^4$ A/m) | $\mu_r=\Delta B/\Delta H$ at ~250 Oe (~2×10$^4$ A/m) | $\mu_r=\Delta B/\Delta H$ at ~500 Oe (~4×10$^4$ A/m) |
| Amumetal-Regular | 8670.28 | 433.1 | 32.67 | 21.7 |
| Amumetal-Annealed | 12640.10 | 452.92 | 31.48 | 20.8 |
| Amumetal-Stressed | 3723.10 | 374.61 | 80.42 | 32.9 |
| A4K-Regular | 16688.68 | 429.44 | 30.84 | 21.4 |
| A4K-Annealed | 51904.52 | 422.83 | 28.552 | 20.7 |
| A4K-Stressed | 10080.62 | 402.91 | 61.97 | 28 |

**Table. 2.** Permeability at 300K

| Material | Permeability ($\mu_r$) at 300K | | | |
|---|---|---|---|---|
| | $\mu_r=\Delta B/\Delta H$ at 0.5 Oe (~40 A/m) | $\mu_r=B/H$ at ~250 Oe (~2×10$^4$ A/m) | $\mu_r=\Delta B/\Delta H$ at ~250 Oe (~2×10$^4$ A/m) | $\mu_r=\Delta B/\Delta H$ at ~500 Oe (~4×10$^4$ A/m) |
| Amumetal-Regular | 10296.49 | 356.00 | 27.50 | 19.5 |
| Amumetal-Annealed | 11662.11 | 356.10 | 27.50 | 19.2 |
| Amumetal-Stressed | 8473.87 | 329.24 | 59.20 | 26.2 |
| A4K-Regular | 4102.02 | 357.19 | 27.90 | 20.0 |
| A4K-Annealed | 11676.87 | 359.41 | 27.50 | 19.3 |
| A4K-Stressed | 2839.67 | 345.32 | 40.30 | 22.4 |

At intermediate fields (~250 Oe) the differential permeability of the stressed samples is better than that of either the annealed or the regular samples. This is because the annealed (and regular) samples tend to almost reach saturation at low fields, thereafter the increase in magnetization with increasing field is small. In contrast, the stressed

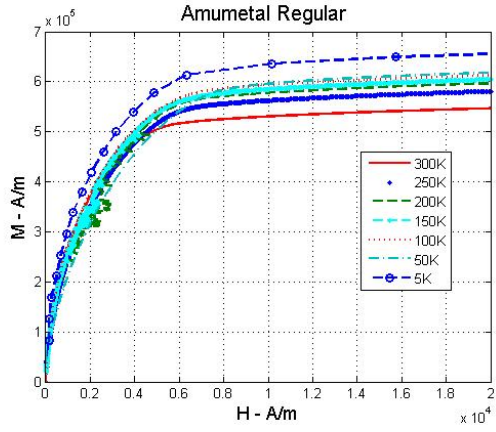

**FIG. 2.** M-H curves for regular amumetal sample at various temperatures.

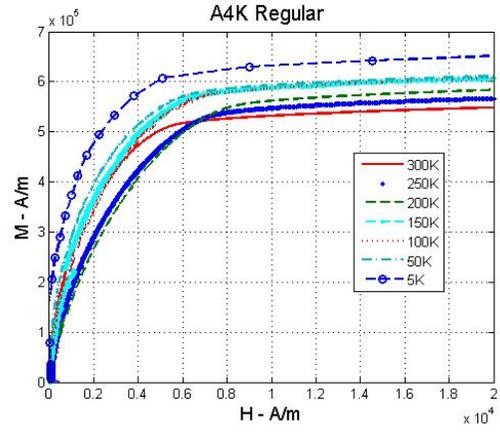

**FIG. 5.** M-H curves for regular A4K sample at various temperatures.

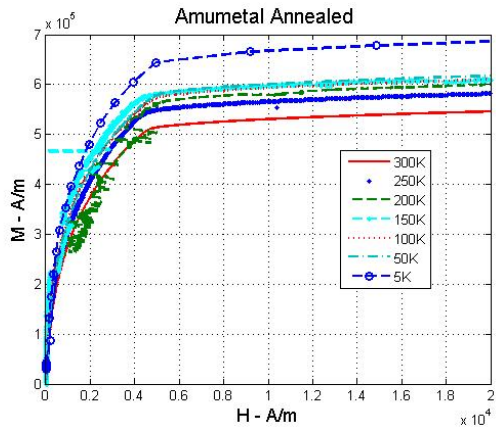

**FIG. 3.** M-H curves for annealed amumetal sample at various temperatures.

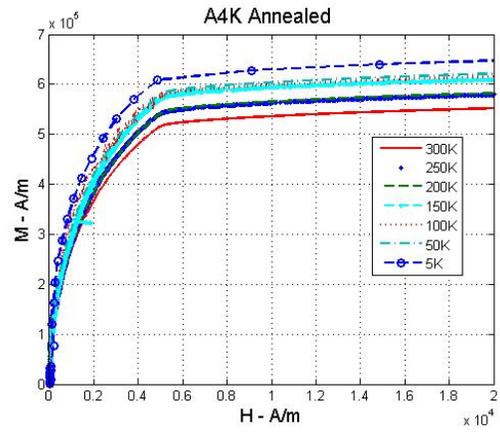

**FIG. 6.** M-H curves for annealed A4K sample at various temperatures.

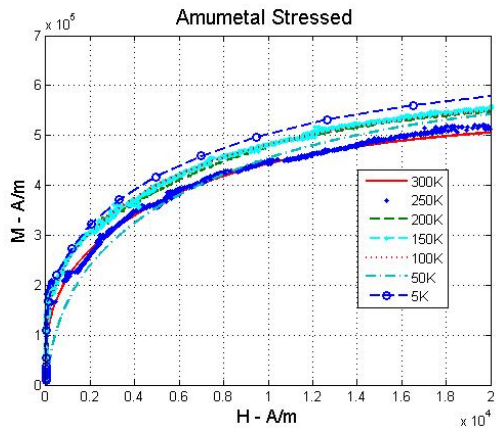

**FIG. 4.** M-H curves for stressed amumetal sample at various temperatures.

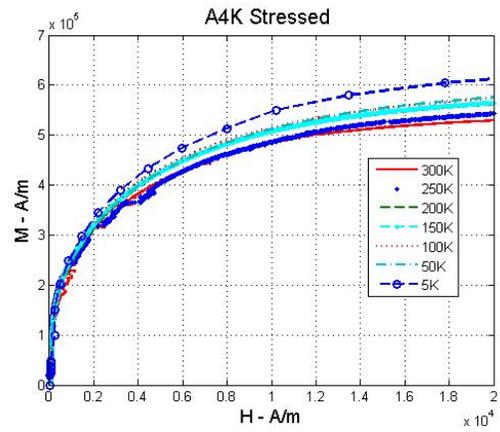

**FIG. 7.** M-H curves for stressed A4K sample at various temperatures.

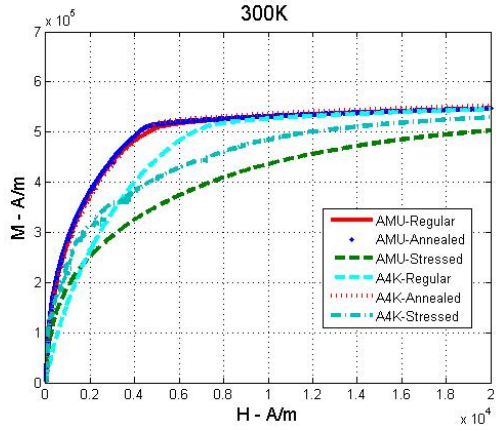

**FIG. 8.** M-H curves for all samples at High field and 300K.

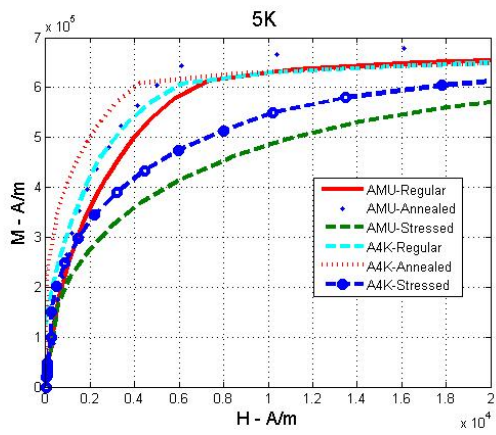

**FIG. 9.** M-H curves for all samples at High field and 5K.

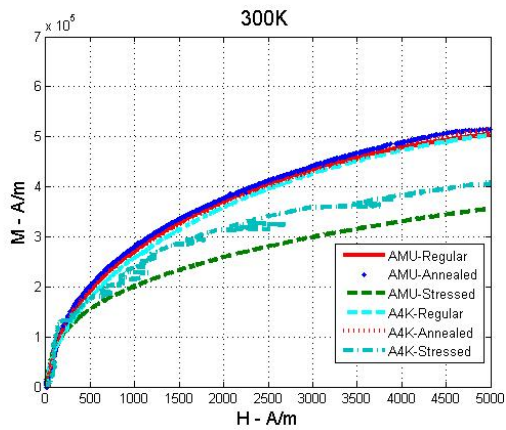

**FIG. 10.** M-H curves for all samples at low field and 300K.

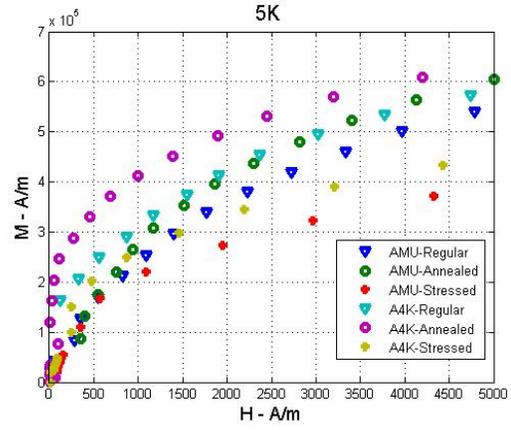

**FIG. 11.** M-H curves for all samples at low field and 5K.

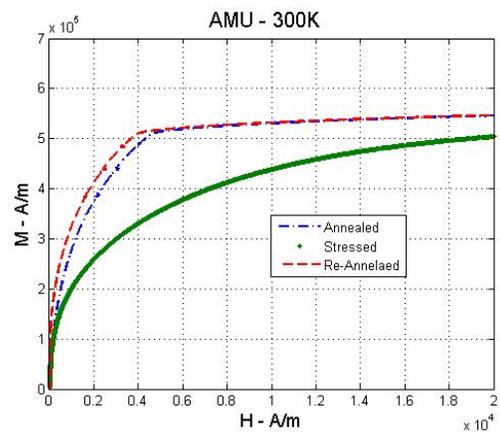

**FIG. 12.** M-H curves for an AMU sample at 300K.

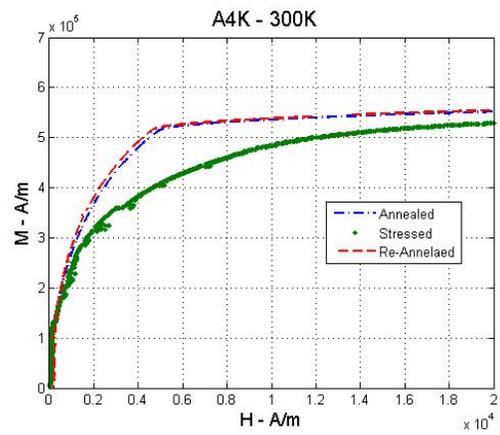

**FIG. 13.** M-H curves for an A4K sample at 300K.

samples need a larger field to drive close to saturation and hence they show a higher differential permeability compared to the annealed (and regular) samples at intermediate fields. This trend is less pronounced at higher field ~500 Oe and it is expected that they would be roughly comparable ($\mu_r \sim 1$) at very higher fields as the magnetization in all samples would reach saturation. However, even at intermediate fields (~250 Oe) if one looks at the absolute permeability (B/H) instead of the differential permeability, at either 0 K or 300K the annealed samples are highest followed by regular and the stressed samples have the least permeability (least B or M for a given H).

We also note that there is some anomalous behavior at the intermediate temperatures 50K-250K in Figures 2-7. Specifically, it appears that is some cases (see for example, Fig 2) the 200K and 250 K appears to have lower permeability at low fields compared to 300K followed by a crossover point as they take higher fields for the M-H fields to nearly "flatten out" compared to the 300 K M-H curves. These trends were found to be repeatable across different samples.

Next, the M-H curves at room temperature (300 K) and cryogenic temperature (5K) are plotted for high fields (Fig 8 and 9) and low fields (Fig. 10 and 11) for both Amumetal and A4K samples. This again shows the permeability decreases greatly due to deformation. This effect is particularly large at 5 K as the thermal energy avoided to overcome pinning defects introduced due to the deformation is very small.

## IV. CONCLUSIONS

An extensive and detailed magnetic characterization of Amumetal and A4K was performed. The results show, deformation due to the manufacturing process has a significant effect on permeability and can be detrimental to magnetic shielding. For the magnetic shielding at room temperature, either annealed Amumetal or annealed A4K can be used as both have relatively comparable permeability. However, annealed A4K has relatively higher permeability at low-field (~0.5 Oe) and low temperature (~5K) and will be more efficient for shielding at these temperatures compared to annealed Amumetal.

Compared to deformed samples, annealed samples of both A4K and Amumetal show a significant improvement in permeability at low fields (~0.5G) at low temperature (5K) compared to its effect at higher temperature (300K). This is possibly due to the fact that at low temperature there is minimal thermal noise to overcome pining defects (abundant in deformed samples) which makes it harder to align the magnetization with a small field compared to annealed samples (fewer pinning sites).

Furthermore, the permeability is more or less restored after the stressed samples are annealed again as shown in figure 12 and figure 13. Since, we found on room temperature magnetic characterization, that the magnetic properties of stressed samples were restored upon annealing, we did not repeat the low temperature magnetic characterization on the stressed samples that were re-annealed as we expect to find that the low temperature magnetic properties will be recovered as well.

## V. ACKNOWLEDGEMENTS

We acknowledge a collaboration between Virginia Commonwealth University (VCU) and Jefferson Lab (U.S. DOE Contract No. DE-AC05-06OR23177) that partly supports Sanjay Sah. We acknowledge Dr. Sama Bilbao Y Leon at VCU Mechanical and Nuclear Engineering for travel support to attend magnetic shielding workshop at FRIB, Dr. Brian Hinderliter at Univ. of Minnesota, Duluth for earlier discussion on Sanjay Sah's PhD research topic, Mr. Michael Adolf at Amuneal Corp. for Amumetal and A4K samples, NCC at VCU for use of the VSM and Prof. Greene and Dr. S. Saha at University of Maryland for use of SQUID Magnetometer.

## VI. REFERENCES

[1] G. Cheng, E. F. Daly, and W. R. Hicks, C100 Cryomodule Magnetic Shielding Finite Element Analysis (Jefferson Lab, 2008), JLAB-TN-08-015.

[2] M. Masuzawa, N. Ohuchi, A. Terashima, and K. Tsuchiya, Applied Superconductivity, IEEE

Transactions Volume:20 , Issue: 3, 1773 - 1776 (2010).

[3] T. Rikitake, Magnetic and Electromagnetic Shielding (Springer, 1987).

[4] R. E. Laxdal, Review of Magnetic Shielding Designs of Low-beta Cryomodules, in SRF2013, Paris, France, (2013).

[5] Material data sets of amumetal and A4K, Amuneal Manufacturing Corporation (private communication).

[6.] R. C. O'Handley, Modern Magnetic Materials Principles and Applications (John Wiley & Sons, Inc., 1999).

[7.] F. Fiorillo, Characterization and Measurement of Magnetic Materials (Elsevier 2004).

[8] O. L. Boothby and R. M. Bozorth, Journal of Applied Physics Volume:18 , Issue: 2 (1947).

[9] A. Zieba and S. Foner, Review of Scientific Instruments Volume:53 , Issue: 9 (1982).

[10] A. Aharoni, Journal of Applied Physics Volume 83, Issue 6, pp. 3432-3434 (1998).

[11] S. Tumanski, Handbook of Magnetic Measurements (CRC Press, 2011), p. 390.

[12] J. Vargasa, C. Ramosa, R. D. Zyslera, and H. Romerob, Physica B: Condensed Matter Volume 320, Issues 1–4 (2002).

[13] K. Suzuki, N. Kataoka, A. Inoue, A. Makino, and T. Masumoto, Materials Transactions, JIM Vol.31 No.8 (1990).

# VIII. APPENDIX

## 1. Demagnetizing Factors

The effective field inside the sample that induces magnetization in the sample is given as: [7]

$$H_{in} = H_{app} - NM, \qquad (3)$$

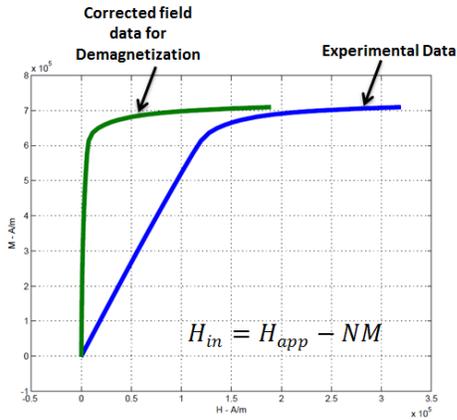

The demagnetizing factor (N) in equation 3 can be written as:

$$N = \frac{H_{app}}{M} - \frac{H_{in}}{M}, \qquad (4)$$

$H_{in}/M$ is the magnetic susceptibility of the material.

$$N = \frac{H_{app}}{M} - \frac{1}{\chi}, \qquad (5)$$

$$\frac{1}{\chi} \approx 0$$

Since, the magnetic susceptibility is very high ($\chi \sim 10{,}000$) for the ferromagnetic material it can be neglected as a first approximation. Hence, the demagnetizing demagnetizing factor (N) can be directly estimated from experimental data as:

$$N = \frac{H_{app}}{M}, \qquad (6)$$

The "N" thus determined was used to correctly estimate the $H_{in}$ using equation 3. All M-H curves plotted in this paper employ this correction to plot M vs. the $H_{in}$, from the measured M vs. $H_{app}$ data.